\documentclass{article}

\usepackage{arxiv}

\usepackage[utf8]{inputenc} 
\usepackage[T1]{fontenc}    
\usepackage{hyperref}       
\usepackage{url}            
\usepackage{booktabs}       
\usepackage{amsfonts}       
\usepackage{nicefrac}       
\usepackage{microtype}      
\usepackage{graphicx}
\usepackage{amsmath}
\usepackage{multirow}
\usepackage[numbers,sort&compress]{natbib}
\usepackage{doi}

\title{LLM-assisted Generation of Pseudo-C2 Servers\\
for IoT Malware Dynamic Analysis}

\author{{Kouki Hasui}\\
	Kagawa University\\
	Takamatsu, Japan\\
	\And
	{Shingo Matsugaya}\\
	Trend Micro, Inc.\\
	Tokyo, Japan\\
 	\And
	{Makoto Shimamura}\\
	Trend Micro, Inc.\\
	Tokyo, Japan\\
 	\And
	\href{https://orcid.org/0000-0001-5596-282X}{\includegraphics[scale=0.06]{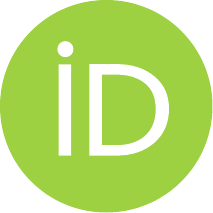}\hspace{1mm}Masaki Hashimoto}\\
	Kagawa University\\
	Takamatsu, Japan\\
}

\renewcommand{\shorttitle}{LLM-assisted Pseudo-C2 Generation for IoT Malware}

\hypersetup{
pdftitle={LLM-assisted Generation of Pseudo-C2 Servers for IoT Malware Dynamic Analysis},
pdfauthor={Kouki Hasui, Shingo Matsugaya, Makoto Shimamura, Masaki Hashimoto},
pdfkeywords={IoT Malware, Mirai, Pseudo-C2 Server, Automatic Generation, Static Analysis, Large Language Model},
}

\begin{document}
\maketitle

\begin{abstract}
Most IoT malware operates as botnets dependent on Command and Control (C2) servers, but the short-lived nature of attack infrastructure often leaves samples dormant without C2 communication, hindering dynamic analysis. This paper proposes a system that combines Ghidra with a Large Language Model (LLM) to extract communication specifications from a malware binary and automatically generate a pseudo-C2 server. Experiments using Mirai demonstrate that the proposed system semantically interprets binary control structures and extracts all 20 core protocol elements in agreement with the ground truth (100\% specification extraction accuracy). The generated pseudo-C2 server fully reproduces seven of ten DDoS attack vectors with attack behavior consistent with the original C2. When applied to a customized variant created by modifying the publicly available Mirai source code, the method succeeds end-to-end---from specification extraction through pseudo-C2 generation to attack reproduction---demonstrating that the LLM infers specifications from binary structures without relying on pre-trained knowledge. This approach extends the applicability of LLMs from analysis assistance to the automated construction of dynamic analysis environments.
\end{abstract}


\section{Introduction}\label{sec:introduction}
In recent years, malware targeting IoT devices has caused increasing damage. According to the NICTER Observation Report 2024 published by the National Institute of Information and Communications Technology (NICT)\cite{NICT2025}, the number of cyberattack-related communications observed in 2024 reached an all-time high, with a substantial portion targeting IoT devices. In particular, since the 2016 source code release of ``Mirai''\cite{Mirai_GitHub}, countless variants have proliferated, making rapid response to the growing volume of samples an urgent challenge.

The analysis of IoT malware faces two major technical barriers. First, IoT malware targets a variety of CPU architectures such as MIPS and ARM, and many samples are stripped of symbol information, demanding considerable expertise and cost for static analysis\cite{Gomes2025}. Second, attackers tend to take down C2 servers within short periods\cite{Alrawi2021}; when the C2 server is no longer operational at the time of analysis, the malware remains dormant and exhibits no attack behavior, making it difficult to identify the threat through dynamic analysis. To address this problem, analysts have traditionally constructed ``observation C2 servers'' by manually reverse-engineering each binary\cite{gdata2020iotbotnet}, but this process incurs high analysis costs and cannot keep pace with the rapidly growing volume of variants.

To overcome these challenges, this paper proposes a system that combines the reverse engineering tool Ghidra with a Large Language Model (LLM) to extract communication specifications from IoT malware binaries and automatically generate a pseudo-C2 server. The contributions of this study are threefold:
\begin{enumerate}
    \item \textbf{Logical extraction of communication specifications by an LLM and demonstration of generality}: We show that an LLM, through static analysis, can comprehensively extract connection parameters, packet structures, and attack command systems even from binaries lacking symbol information. In particular, by logically interpreting binary control structures (such as \texttt{switch-case} statements), the LLM accurately identifies even ``missing entries'' in the attack vector ID system without hallucination, demonstrating analytical capabilities that go beyond simple keyword extraction. In the evaluation on Mirai, the LLM extracted all \textbf{20 core elements} of the communication protocol in agreement with the ground truth. Furthermore, when applied to a customized variant created by intentionally modifying the publicly available Mirai source code, the proposed method correctly extracted the modified values, confirming that the LLM infers specifications from binary structures without relying on pre-trained knowledge and thereby establishing the generality of the approach.
    
    \item \textbf{Reproduction of attack behavior}: By issuing diverse attack commands from the automatically generated pseudo-C2 server, the system fully reproduced 7 of 10 attack vectors with protocol-conformant attack packets matching the behavior under the original C2.
    
    \item \textbf{Automated construction of a dynamic analysis platform}: Through these capabilities, the proposed system constructs a dynamic analysis environment that does not depend on external C2 servers, automating the majority of an analysis process that traditionally required advanced reverse engineering skills.
\end{enumerate}

The remainder of this paper is organized as follows. Section~\ref{sec:related} reviews related work, and Section~\ref{sec:method} describes the proposed method in detail. Section~\ref{sec:experiment} presents the experimental design and results, and Section~\ref{sec:discussion} discusses the findings. Finally, Section~\ref{sec:conclusion} concludes the paper.

\section{Related Work}\label{sec:related}
This section organizes existing approaches to dynamic analysis of IoT malware along the axis of ``C2 server dependence,'' and clarifies the positioning of the present study.

\subsection{Challenges in Dynamic Analysis of IoT Malware}
To comprehensively grasp the impact that malware ultimately has on a system, dynamic analysis---in which the sample is executed in an isolated environment and its communications and system calls are observed---is indispensable. However, much IoT malware is of the botnet type that begins activity only upon receiving commands from a C2 server, and this introduces an inherent challenge: unless communication with the C2 server is established, the malware does not exhibit its intended malicious behavior.

In response to this challenge, research on dynamic analysis platforms specialized for IoT environments has been advanced. ``V-Sandbox,'' proposed by Le et al.\cite{Le2020}, incorporates countermeasures against IoT-malware-specific anti-analysis techniques into a QEMU-based emulation environment, enabling the observation of concealed behavior. However, while such sandboxes resolve the execution-environment challenges of running malware on heterogeneous architectures, the reception of attack commands still requires a live, operational C2 server, leaving the fundamental issue of C2 dependence unresolved.

\subsection{Approaches to C2-Independent Analysis and Their Limitations}
To enable dynamic analysis even for samples whose live C2 servers have been taken down, four broad approaches have been attempted.

The first approach actively probes the Internet for live C2 servers. ``CnCHunter,'' by Davanian et al.\cite{Davanian2021}, identifies C2 infrastructure using a man-in-the-middle (MITM) technique, and its successor ``C2Miner''\cite{Davanian2024_C2Miner} uses old malware samples as decoys to discover live C2 servers. However, since their success depends entirely on the existence of a live C2 server, they cannot be applied in the modern threat landscape, where attack infrastructures are short-lived\cite{Alrawi2021}, to samples whose C2 has already been taken down.

The second approach involves analysts manually reverse-engineering the binary and constructing a local observation C2 server\cite{gdata2020iotbotnet}. While this approach is independent of C2 availability, it requires advanced reverse engineering skills and substantial human cost for each sample, making rapid response to the daily emergence of new variants difficult.

The third approach attempts to automatically generate counterpart systems. Musch et al.\cite{Musch2018_Chameleon} proposed ``Chameleon,'' which targets web applications and constructs a low-interaction honeypot by learning response templates from observations of real systems; however, this presupposes a live system as the learning source and cannot be directly applied to samples whose C2 servers have been taken down. Borzacchiello et al.\cite{Borzacchiello2019} applied symbolic execution to remote access trojan (RAT) binaries to synthesize pseudo-C2 servers, but symbolic execution faces critical weaknesses such as path explosion and the difficulty of constraint solving for cryptographic processing, and its target is limited to RATs, posing scalability limitations for modern IoT malware with custom loop processing and obfuscation.

The fourth approach reactively generates counterpart communication while observing malware responses during dynamic analysis. ``RIoTMAN,''\cite{Darki2020_RIoTMAN} proposed by Darki et al., achieves C2-server impersonation through iterative adaptation of the target environment configuration and automated engagement with the malware, demonstrating that approximately 79\% of 3{,}024 IoT malware samples could be induced to transition to DDoS-attack or proliferation phases. While RIoTMAN is the prior study most closely aligned with the present work in problem awareness, its reactive response generation does not entail deep protocol understanding of the binary, leading to inherent limitations in the comprehensiveness of communication specifications and in the qualitative fidelity of attack reproduction.

\subsection{Application of LLMs to Malware Analysis}
In recent years, research applying the high inferential capabilities of Large Language Models (LLMs) to malware analysis has been accelerating\cite{Jelodar2025}. LLMs can infer the original semantics of processing from instruction patterns and context even in assembly code that lacks symbol information, and have the advantage of avoiding path explosion because they do not depend on rigorous mathematical models like symbolic execution.

As representative studies, ``IRIS,'' by Li et al.\cite{Li2025}, automatically generates analysis rules (taint specifications) for a static analysis tool (CodeQL) using an LLM, improving vulnerability detection capabilities by approximately twofold. Fujii and Yamagishi\cite{Fujii2024} input decompiled malware code into an LLM to generate functional descriptions, and demonstrated that they could comprehensively explain malware behavior with up to 90.9\% accuracy.

However, these studies position the LLM merely as an ``auxiliary tool for static analysis,'' and there is no example that goes as far as automatically constructing dynamic-analysis infrastructure (such as a pseudo-C2 server) capable of actually interacting with malware based on the extracted specifications.

\subsection{Positioning of This Study}
A comparison between the related work described above and the present study is organized in Table~\ref{tab:research_positioning}.

\begin{table}[t]
  \caption{Comparison of related work with this study}
  \label{tab:research_positioning}
  \centering
  \scriptsize
  \setlength{\tabcolsep}{4pt}
  \begin{tabular}{p{2.4cm} p{3.6cm} c c p{4.0cm}}
    \hline
    \textbf{Category} & \textbf{Representative work} & \textbf{LLM} & \textbf{C2 dep.} & \textbf{Limitations / Difference from this work} \\
    \hline \hline
    Dynamic analysis\newline (platform/observation) & Le \cite{Le2020} (V-Sandbox) & $\times$ & Required & A live C2 is required to receive attack commands \\
    \hline
    Dynamic analysis\newline (C2 discovery) & Davanian \cite{Davanian2021, Davanian2024_C2Miner} & $\times$ & Mandatory & Analysis becomes infeasible after C2 takedown \\
    \hline
    Pseudo-C2 construction\newline (static/automatic) & G DATA \cite{gdata2020iotbotnet}, Musch \cite{Musch2018_Chameleon}, Borzacchiello \cite{Borzacchiello2019} & $\times$ & Not required & High human cost / requires learning source / path explosion / RAT-only \\
    \hline
    Pseudo-C2 construction\newline (dynamic/reactive) & Darki \cite{Darki2020_RIoTMAN} (RIoTMAN) & $\times$ & Not required & Reactive response generation limits comprehensiveness of specifications \\
    \hline
    LLM application\newline (analysis support) & Li \cite{Li2025} (IRIS), Fujii \cite{Fujii2024} & $\bigcirc$ & Not required & Limited to static analysis support; does not reach attack-infrastructure generation \\
    \hline
    \textbf{This work} & \textbf{Pseudo-C2 automatic generation} & \textbf{$\bigcirc$} & \textbf{Not required} & \textbf{Automatically generates a C2 based on LLM-driven specification extraction; full observation possible even after C2 takedown} \\
    \hline
  \end{tabular}
\end{table}

The novelty of this study lies in applying the inferential capabilities of LLMs not only to ``analysis'' but also to the ``automatic generation of attack infrastructure (a pseudo-C2 server).'' Specifically, based on specifications extracted by static analysis, the LLM autonomously generates implementation code for a C2 server. This makes it possible to forcibly establish communication in an isolated environment and to observe detailed attack behavior even for samples whose live C2 servers have been taken down. In other words, the present method is an approach that simultaneously realizes ``automation of static analysis'' and ``elimination of C2 dependence in dynamic analysis,'' positioning it as an integrated analysis platform that comprehensively addresses the limitations of existing research.

\section{Proposed Method}\label{sec:method}

\subsection{System Overview}

\begin{figure}[t]
\begin{center}
\includegraphics[keepaspectratio, width=0.75\linewidth]{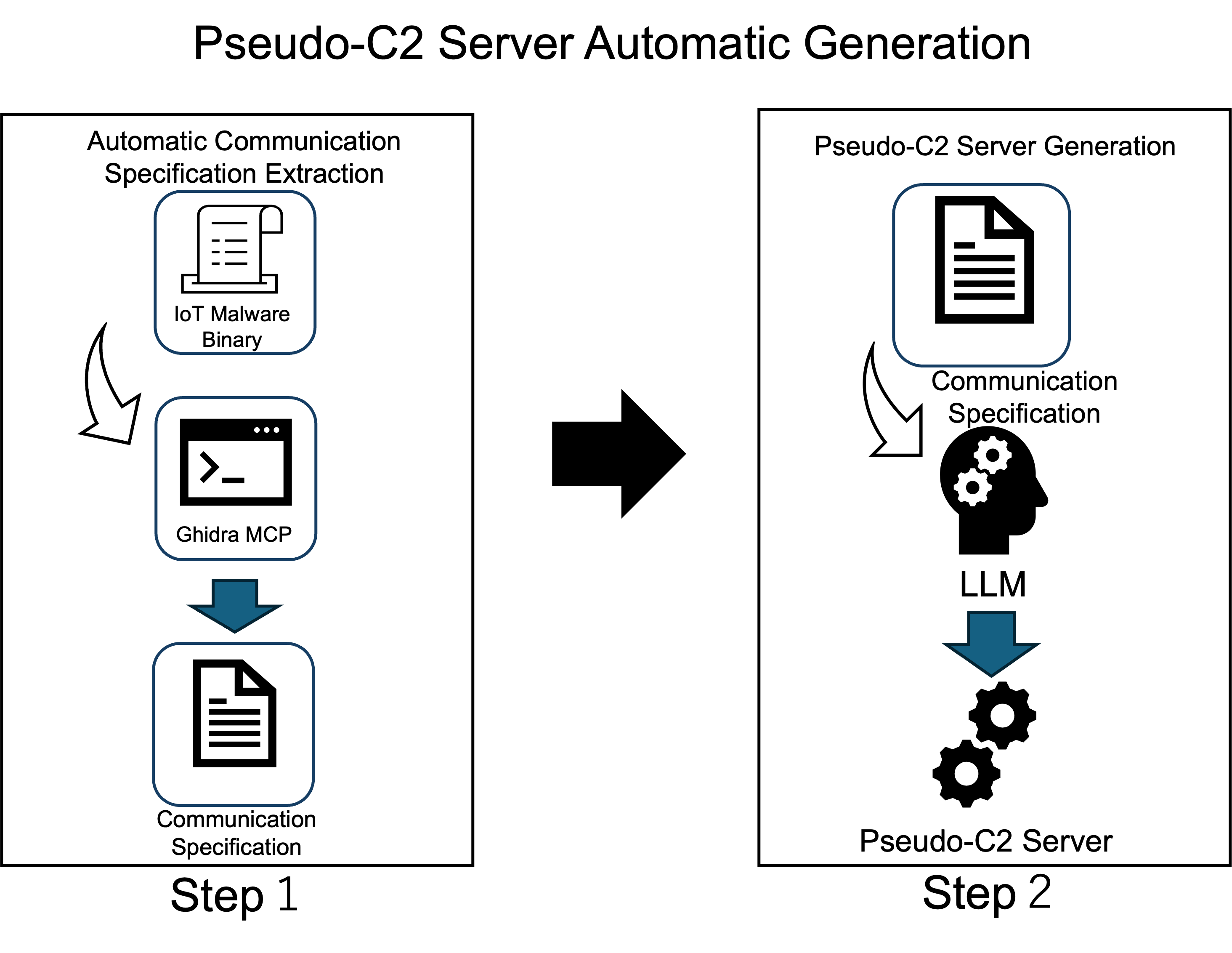}
\caption{Overall flow of the proposed pseudo-C2 server automatic generation}
\label{fig:system_flow}
\end{center}
\end{figure}

This study proposes a ``pseudo-C2 server automatic generation system.'' As illustrated in Figure~\ref{fig:system_flow}, the proposed system consists of the following two steps.

\begin{enumerate}
    \item \textbf{Communication specification extraction (Step 1)}: Taking an IoT malware binary as input, the system extracts the protocol and command system used to communicate with the C2 server through static analysis using Ghidra and an LLM, and outputs the result as a textual specification document.
    \item \textbf{Pseudo-C2 server generation (Step 2)}: Using the specification document obtained in Step 1 as input, the system feeds it to the LLM to automatically generate a pseudo-C2 server program that can communicate with the malware.
\end{enumerate}

By providing the specification document obtained in Step 1 as a clear constraint to the LLM in Step 2, this two-step design enables the malware analyst to construct a dynamic analysis environment without depending on advanced reverse engineering skills.

\subsection{Communication Specification Extraction (Step 1)}\label{subsec:step1}

In Step 1, the malware binary is loaded into Ghidra, and the LLM analyzes the decompiled output to extract the communication specifications required for connecting to the C2 server. The information items targeted for extraction are organized into three categories as shown in Table~\ref{tab:spec_format}: connection establishment, protocol structure, and command system. The LLM infers each of these items from constant definitions, control structures (such as conditional branches and \texttt{switch-case} statements), and API calls within the decompiled pseudo-code, and outputs the result as a textual specification document.

\begin{table}[t]
\caption{Structure of the communication specification document extracted in Step 1}
\label{tab:spec_format}
\centering
\footnotesize
\begin{tabular}{p{2.8cm}p{3.0cm}p{6.5cm}}
\hline
\textbf{Category} & \textbf{Item} & \textbf{Content} \\
\hline \hline
\multirow{2}{*}{Connection} & C2 server address & The destination to which the malware attempts to connect \\
 & Handshake data & Verification byte sequences (e.g., magic numbers) used to authenticate the communication \\
\hline
\multirow{2}{*}{Protocol structure} & Header/payload structure & Header size and the position/size of the payload length field \\
 & Endianness & Byte order of numerical data (big/little) \\
\hline
\multirow{2}{*}{Command system} & Attack vector ID & Mapping between attack command IDs from the C2 and the corresponding attack methods (e.g., SYN flood) \\
 & Argument structure & Serialization format of attack targets, durations, etc. \\
\hline
\end{tabular}
\end{table}

In particular, since most IoT malware obfuscates its C2 server location and related strings using custom encryption routines to conceal the C2 address, the LLM analyzes such obfuscation routines (e.g., table-based XOR encryption) to penetrate the obfuscation layer and extract the correct values. This capability enables the proposed method to reach analytical depths unattainable by simple static string extraction, distinguishing it from conventional signature-based analysis.

\subsection{Tool Support Mechanism Design}\label{subsec:tool_support}

The proposed system adopts the publicly available tool ``GhidraMCP''\cite{LaurieWired} as the foundation for providing binary analysis context from Ghidra to the LLM. However, through the evaluation experiments in this study, we found that the existing GhidraMCP feature set occasionally fails to provide the LLM with the information needed for certain analysis targets. Specifically, the existing implementation lacks adequate functionality for retrieving constant values hardcoded at specific binary addresses (such as magic numbers and encryption keys) and short string constants referenced after decryption. While GhidraMCP supports the retrieval of high-level information such as decompilation results and call graphs, it does not provide the low-level functionality of directly accessing raw byte sequences stored at specific memory addresses. As a result, the LLM cannot receive such information as context, leading to extraction failures (details are described in Section~\ref{subsec:experiment2}).

To address this limitation, we extended GhidraMCP by implementing three additional MCP tools: (i) \texttt{read\_bytes}, which reads raw byte sequences from arbitrary memory addresses in hexdump format; (ii) \texttt{read\_string\_at}, which reads NULL-terminated ASCII strings; and (iii) \texttt{get\_section\_info}, which retrieves the list of memory blocks (\texttt{.text}, \texttt{.rodata}, \texttt{.data}, \texttt{.bss}, etc.) of an ELF binary. These extensions enable the LLM to access low-level information stored at arbitrary memory addresses within the binary, including raw byte sequences, NULL-terminated strings, and memory block layouts. As demonstrated in Section~\ref{subsec:experiment2}, this tool extension plays an essential role in eliciting the inferential capabilities of the LLM and enabling specification extraction based on binary structures rather than pre-trained knowledge. This tool extension is positioned as an essential component of LLM-assisted binary analysis.

\subsection{Pseudo-C2 Server Generation (Step 2)}

In Step 2, the communication specification document output in Step 1 is fed to the LLM as a prompt to automatically generate a pseudo-C2 server capable of communicating with the malware. Python is specified as the implementation language, given its ease of socket communication control and high readability.

The prompt design for the generation process emphasizes three points. First, the contents of the specification document are explicitly stated as constraints to be strictly followed, ensuring binary compatibility for low-level specifications such as endianness and header length. Second, detailed log output and error handling are required to facilitate troubleshooting when communication fails. Third, a natural-language interface for issuing attack commands is required to enhance analyst usability.

The pseudo-C2 server generated as a result possesses the following capabilities to support dynamic analysis:
\begin{itemize}
    \item \textbf{Handshake establishment}: Verifies post-connection data (such as magic numbers) and maintains communication only with legitimate bots, thereby excluding unrelated scanning traffic.
    \item \textbf{Attack command generation}: Automatically converts natural-language attack names entered by the analyst (e.g., ``SYN flood'') into the malware-specific binary values (Vector IDs) based on the ID correspondence table extracted in Step 1, allowing the analyst to invoke arbitrary attack methods through intuitive operations without needing to be aware of the binary values.
    \item \textbf{Debugging features}: Logs all sent and received packets in real time in both hexadecimal (hexdump) and ASCII format, enabling byte-level verification of consistency between the specifications and actual communication packets.
\end{itemize}

\section{Experimental Design and Results}\label{sec:experiment}

\subsection{Experimental Design}

\subsubsection{Target and Evaluation Approach}
To verify the effectiveness of the proposed method, we conduct empirical experiments targeting the publicly available binary of ``Mirai,'' a representative IoT malware whose specifications are known. As the reference dataset, we adopt ``Mirai-Source-Code,''\cite{Mirai_GitHub} a repository published on GitHub by J. Gamblin. This repository is curated from the leaked Mirai source code with incomplete files and configurations removed to allow compilation, and is widely referenced as the standard reference implementation of Mirai in the security research field.

The rationale for selecting Mirai as the verification target is threefold. First, since the leak in September 2016, Mirai has continued to spawn numerous variants such as ``Satori'' and ``Okiru,''\cite{Affinito2023} establishing it as the de facto standard of the IoT threat landscape. Second, because the communication protocol definitions described in the public source code (under \texttt{mirai/bot/}) are available, the specifications extracted by the proposed method from binaries can be rigorously compared against the known ground truth. Third, Mirai exhibits a structure in which a wide range of DDoS attack vectors over UDP/TCP/GRE/HTTP/DNS are controlled by a single-byte attack ID (Vector ID), making it well-suited for verifying the comprehensiveness of specification extraction.

\subsubsection{Evaluation Metrics}
We evaluate the effectiveness of the proposed method using the following two metrics.

\textbf{(1) Specification Extraction Accuracy}: The percentage of items, among the core elements of the Mirai communication protocol (connection parameters and attack vector ID system) in the specification document obtained in Step 1, that match the ground truth derived from the public source code.
\begin{equation}
    \text{Extraction Accuracy (\%)} = \frac{\text{Number of items matching the ground truth}}{\text{Total number of ground-truth items}} \times 100
\end{equation}

\textbf{(2) Attack Reproduction Success Rate}: The classification of each attack method into the following three categories based on the result of issuing attack commands from the pseudo-C2 server generated in Step 2.
\begin{itemize}
    \item \textbf{Fully reproduced}: The packet structure (protocol type, flags, header, payload length, etc.) matches that of the original C2, and the attack is carried out to completion.
    \item \textbf{Partially reproduced}: The bot receives the attack command and starts the attack function, but the attack is not completed due to environmental factors.
    \item \textbf{Environmental constraint}: The attack behavior itself cannot be observed due to the constraints of the closed experimental environment.
\end{itemize}

\subsubsection{Network Configuration}
The experimental environment was constructed on a closed network built on a virtualization platform, comprising three virtual machines: a Bot machine (\texttt{192.168.1.102}) that executes the Mirai samples, a C2 machine (\texttt{192.168.1.10}) that runs either the original C2 or the pseudo-C2 server, and a Target machine (\texttt{192.168.1.50}) that observes incoming attack packets via \texttt{tcpdump}. The network is physically and logically isolated from the external Internet, eliminating the risk of malware leaking outside. Considering the ease of dynamic analysis, we executed Mirai samples on the bot machine that had been cross-compiled to x86 32-bit, taking into account that the majority of target IoT devices use 32-bit processors and following the standard build configuration of the reference Mirai-Source-Code repository. This applies to both Experiment I and Experiment II.

In addition, we individually verified the completion of the measurement interval (10 seconds) for every attack method through packet captures, confirming that no measurement was interrupted or truncated.

\subsubsection{Setup of Communication Control}
The Mirai sample used in this experiment attempts to connect to a hard-coded destination IP (\texttt{65.222.202.53}) and TCP port 80, and additionally judges whether to continue running based on whether Google Public DNS (\texttt{8.8.8.8}) responds at startup. To establish communication with the pseudo-C2 in a closed environment without modifying the binary at all, we applied a three-stage communication control to the bot machine at the OS level using \texttt{iptables} and routing tables.

First, a static route to the hard-coded destination was added, which prevents immediate packet drops due to ``unknown destination'' in an environment without a default gateway and allows the sample's send operations to complete normally. Second, traffic to \texttt{8.8.8.8} was redirected via DNAT to the local pseudo-C2, returning a pseudo response that satisfies the malware's Internet-connectivity check and thereby allowing the bot's startup sequence to proceed. Third, packets destined for the hard-coded C2 address (\texttt{65.222.202.53:80}) were transparently forwarded via DNAT and port translation to the pseudo-C2 server (\texttt{192.168.1.10:23}).

These controls realize an environment in which the sample can be operated under the control of the pseudo-C2 server without any modification to the sample, even in the closed environment. Note that in Experiment II described in Section~\ref{subsec:experiment2}, the C2 connection destination on the sample side is intentionally changed, so the third control is replaced with a setting specific to Experiment II (details are given in Section~\ref{subsec:experiment2}).

\subsubsection{LLM Selection}\label{subsubsec:llm_selection}
For the selection of an LLM, we first conducted preliminary verification using locally executable open-source models (Llama 3 70B, DeepSeek-Coder-V2, etc.) from the perspective of confidentiality. However, in the task of comprehensively extracting C2 communication specifications from the fragmentary decompiled code that Ghidra outputs, these models proved insufficient in contextual understanding, with frequent hallucinations occurring in parameter identification. Accordingly, this study adopts Anthropic's commercial models, which possess higher inferential capabilities, via the CLI interface ``Claude Code.'' Specifically, ``Claude Opus 4.5'' was used in Experiment I (Section~\ref{subsec:experiment1}), and ``Claude Opus 4.7'' was used in Experiment II (Section~\ref{subsec:experiment2}) due to a model update during the experimental period; we confirmed that both versions reliably perform specification extraction and pseudo-C2 server generation under the proposed method. The analysis prompt employs neutral wording such as ``extract the specifications related to the communication of the binary currently selected,'' and no interruption of analysis by the LLM's safety mechanisms occurred.

\subsection{Experiment I: Evaluation on the Original Mirai}\label{subsec:experiment1}

\subsubsection{Communication Specification Extraction Results}\label{subsubsec:exp1_extraction}
As a result of LLM-based analysis, the proposed method successfully extracted the parameters required for C2 communication in a comprehensive manner from the obfuscated initialization function (\texttt{table\_init}) and the attack processing loop (\texttt{attack\_init}) within the binary. The extraction results are shown in Table~\ref{tab:extraction_results}. The targets of extraction comprise four connection parameters (C2 IP address, port, Magic Number, and Bot ID Length), five command-packet structure items (offset, size, and byte order of each field), and 11 attack vector ID items (\texttt{0x00}--\texttt{0x0A}, including the missing entry \texttt{0x08}), totaling 20 items. In particular, all of the following matched the ground truth: the XOR-encrypted C2 IP address, the packet structure recovered by tracing the control flow of the parsing routine (\texttt{attack\_parse}) within the binary, and the Vector ID mapping extracted from the registration logic. A particularly noteworthy point is that the LLM correctly recognized the fact that Vector ID \texttt{0x08} is, by definition, a missing entry; the significance of this observation is discussed in Section~\ref{sec:discussion}.

\begin{table}[t]
\caption{Communication specification extraction results in Experiment I (all 20 items)}
\label{tab:extraction_results}
\centering
\footnotesize
\begin{tabular}{llll}
\hline
\textbf{Category} & \textbf{Item} & \textbf{LLM Extraction Result} & \textbf{Ground Truth} \\
\hline \hline
\multirow{4}{*}{Connection parameters} & C2 IP Address & 65.222.202.53 & 65.222.202.53 \\
 & C2 Port & 80/TCP & 80/TCP \\
 & Magic Number & 0x00000001 & 0x00000001 \\
 & Bot ID Length & 1 byte & 1 byte \\
\hline
\multirow{5}{*}{Packet structure} & Offset 0x00 & Packet Length (2B, BE) & Match \\
 & Offset 0x02 & Attack Duration (4B, NBO) & Match \\
 & Offset 0x06 & Attack Vector ID (1B) & Match \\
 & Offset 0x07 & Number of Targets (1B) & Match \\
 & Offset 0x08-- & Target Entries (IP+mask) & Match \\
\hline
\multirow{11}{*}{Attack Vector ID} & 0x00 & UDP Generic Flood & Match \\
 & 0x01 & UDP VSE Flood & Match \\
 & 0x02 & UDP DNS Flood & Match \\
 & 0x03 & TCP SYN Flood & Match \\
 & 0x04 & TCP ACK Flood & Match \\
 & 0x05 & TCP STOMP Flood & Match \\
 & 0x06 & GRE IP Flood & Match \\
 & 0x07 & GRE Ethernet Flood & Match \\
 & \textbf{0x08} & \textbf{(missing)} & \textbf{Recognized as missing} \\
 & 0x09 & UDP Plain Flood & Match \\
 & 0x0A & HTTP Flood & Match \\
\hline
\end{tabular}
\end{table}

Based on these results, the specification extraction accuracy in this experiment matched the ground truth on all \textbf{20 items}, achieving \textbf{20/20 (100\%)}. Furthermore, the keep-alive mechanism (a 2-byte \texttt{0x0000} sent at 60-second intervals) and other control mechanisms necessary for maintaining a long-lived connection were also extracted, providing the foundational information needed for long-term operation of the pseudo-C2 server.

\subsubsection{Attack Reproduction Results}\label{subsubsec:exp1_attacks}
We verified the attack behavior triggered by the proposed pseudo-C2 server in comparison with the original C2 server. After the bot machine started up, the pseudo-C2 server completely reproduced Mirai's specific handshake sequence and was successfully recognized as a legitimate C2: immediately after TCP connection establishment, it correctly received the magic number \texttt{0x00000001} sent by the bot, completed the bot ID exchange, and then transitioned to the keep-alive maintenance phase, indicating that the LLM accurately understood Mirai's communication initialization sequence.

For TCP-based attacks (SYN flood, ACK flood, STOMP flood), the pseudo-C2 server issued attack commands and the bot generated attack packets accordingly. For SYN/ACK floods, packet structures (sequence numbers, source IPs, destination ports, etc.) consistent with the original C2 environment were observed. For STOMP, although TCP three-way handshake attempts were observed, full attack execution was not completed; this is treated as ``partial reproduction'' on the bot side. For UDP-based attacks (UDP Generic, VSE, Plain), the pseudo-C2 server faithfully reproduced UDP packet generation; in particular, for UDP VSE flood, the bot generated attack packets containing the Source Engine query payload (\texttt{0xFF\,0xFF\,0xFF\,0xFF\,0x54\,Source Engine Query\,0x00}), demonstrating that the LLM accurately interpreted vendor-specific protocol structures from the binary as well. For GRE-based attacks (GRE IP, GRE Ethernet), the pseudo-C2 server triggered the generation of GRE-encapsulated packets; particularly noteworthy is that for GRE Ethernet flood, GRE-encapsulated Ethernet frames were correctly generated, indicating that the LLM was able to handle the unique encapsulation specifications of the GRE protocol.

For DNS Flood and HTTP Flood, no attack packets were observed at the target. Since these attacks presuppose protocol-conformant communication with an external DNS server or web server, the cause is considered to be the absence of the corresponding attack-target services on the closed network. This observation failure is not a limitation of the proposed method but stems from the constraints of the experimental environment, and the issuance of attack commands itself was processed normally on the C2 side. This perspective is discussed in Section~\ref{sec:discussion}.

Table~\ref{tab:reproduction_summary} summarizes the reproduction result for all 10 attack methods. Among the 10 attack vectors, 7 were fully reproduced with protocol-conformant attack packets matching the behavior under the original C2, 1 (TCP STOMP) was partially reproduced (attack command received and connection attempted, but full execution not completed due to dependence on target-side responses), and 2 (DNS Flood, HTTP Flood) failed to be observed due to environmental constraints in the closed network.

\begin{table}[t]
\caption{Reproduction results for all attack methods}
\label{tab:reproduction_summary}
\centering
\footnotesize
\begin{tabular}{ll}
\hline
\textbf{Attack Method} & \textbf{Result} \\
\hline \hline
TCP SYN Flood   & Fully reproduced \\
TCP ACK Flood   & Fully reproduced \\
TCP STOMP Flood & Partially reproduced (connection attempts only) \\
UDP Generic Flood & Fully reproduced \\
UDP VSE Flood   & Fully reproduced \\
UDP Plain Flood & Fully reproduced \\
GRE IP Flood    & Fully reproduced \\
GRE Ethernet Flood & Fully reproduced \\
DNS Flood       & Observation failed (no DNS server in the environment) \\
HTTP Flood      & Observation failed (no web server in the environment) \\
\hline
\end{tabular}
\end{table}

\subsection{Experiment II: Verification with a Customized Variant}\label{subsec:experiment2}

\subsubsection{Purpose of Verification and Design of the Customized Variant}\label{subsubsec:exp2_design}

The results of Experiment I in Section~\ref{subsec:experiment1} demonstrated that the proposed method extracts Mirai's communication specifications with high accuracy (specification extraction accuracy of 100\%). However, since Mirai's source code has been publicly available together with related analysis reports since the leak of September 2016\cite{Mirai_GitHub}, the possibility that the LLM's pre-trained data contains the ground-truth information cannot be ruled out. Accordingly, it is essential to verify whether the results of Experiment I stem from the LLM's pure inferential capabilities based on binary structures or from supplementation by pre-trained knowledge.

To address this verification objective, this section creates a customized variant (hereafter, ``Variant II'') by intentionally modifying the publicly available Mirai source code, and applies the proposed method. The modifications applied to Variant II, together with the LLM extraction results obtained by the proposed method, are shown in Table~\ref{tab:custom_results}. None of these modified values appear in the publicly available Mirai source code, so if the LLM correctly extracts them, this can only be attributed to inference based on the binary structure rather than reliance on pre-trained data.

\begin{table}[t]
\caption{Modifications and extraction results for Variant II}
\label{tab:custom_results}
\centering
\footnotesize
\begin{tabular}{llll}
\hline
\textbf{Item} & \textbf{Original Value} & \textbf{Variant II Value} & \textbf{LLM Extraction Result} \\
\hline \hline
Magic Number & \texttt{0x00000001} & \texttt{0xCAFEBABE} & \texttt{0xCAFEBABE} \\
Encryption key & (original key) & \texttt{0xDEADBEEF} & \texttt{0xDEADBEEF} \\
C2 IP address & \texttt{65.222.202.53} & \texttt{192.168.99.99} & \texttt{192.168.99.99} \\
C2 Port & 80/TCP & 80/TCP & 80/TCP \\
Attack Vector IDs (10) & \texttt{0x00}--\texttt{0x0A} & Randomly reassigned & Exact match with modified values \\
\hline
\end{tabular}
\end{table}

In particular, for the attack Vector IDs, the identifiers for all 10 attack methods were reassigned to random values unrelated to the originals (\texttt{0x91}, \texttt{0x73}, \texttt{0x2d}, \texttt{0x19}, \texttt{0x5c}, \texttt{0x4b}, \texttt{0x82}, \texttt{0x27}, \texttt{0x6a}, \texttt{0x3e}). This was designed such that the LLM cannot reconstruct the ID system unless it correctly analyzes the registration logic of the \texttt{attack\_init} function.

\subsubsection{Communication Specification Extraction Results}\label{subsubsec:exp2_extraction}

For Variant II, we launched a new Claude Code session independent of the one used in Experiment I, and applied Step 1 of the proposed method. This was done to eliminate any influence on the analysis results that residual session context from Experiment I might cause.

As a result, as shown in Table~\ref{tab:custom_results} (in the previous subsection), the LLM successfully extracted values matching the ground truth for Variant II for all of the modified items.

A particularly notable point is that, despite the fact that the attack Vector ID system in Variant II differs significantly from the original (e.g., the value corresponding to TCP SYN Flood, which is \texttt{0x03} in the original, is reassigned to \texttt{0x19} in Variant II), the LLM completely and correctly extracted the new ID system by analyzing the registration logic of the \texttt{attack\_init} function within the binary. This is strong evidence that the LLM is reconstructing specifications from the actual binary structure, without being misled by the legacy system (\texttt{0x00}--\texttt{0x0A}) contained in its pre-trained data.

However, two issues in the experimental design were observed during this verification process. First, regarding the extraction of the C2 IP address, in the initial implementation the old IP (\texttt{65.222.202.53}) was left within the binary in encrypted form, and the LLM accurately decrypted it and consequently extracted the old IP. This was not a problem on the LLM side, but rather originated from incomplete modification of one of the references in the source code (\texttt{mirai/bot/includes.h}); it was resolved by completing the modification. Second, regarding the extraction of the Magic Number, the GhidraMCP we used did not implement the function for retrieving raw byte sequences at specific addresses, so contextual information itself was not provided to the LLM, and the extraction did not succeed. This was addressed through the GhidraMCP extension described in Section~\ref{subsec:tool_support} (Proposed Method). Through subsequent reanalysis after addressing these issues, accurate extraction was achieved for all items. These observations suggest that, in order to elicit the LLM's inferential capabilities, the information-providing capability of the tooling side is decisively important, and we discuss this further in Section~\ref{subsec:tool_importance}.

\subsubsection{Pseudo-C2 Server Generation and Attack Verification}\label{subsubsec:exp2_attacks}

Using the communication specification document extracted from Variant II as input, we automatically generated a pseudo-C2 server using Step 2 of the proposed method. The generated code consists of six files (\texttt{main.py}, \texttt{server.py}, \texttt{protocol.py}, \texttt{attacks.py}, \texttt{repl.py}, \texttt{test\_smoke.py}) and accompanying documentation (\texttt{README.md}), and includes builder functions corresponding to all 10 attack vectors as well as 19 test cases, demonstrating a high-quality implementation. The generated code also includes defensive implementations such as isolated-environment checks, KEEPALIVE handling, and payload-length validation, demonstrating that the LLM autonomously implements comprehensive functionality based on the specification document.

The generated pseudo-C2 server was operated against Variant II, and 10 types of attacks were triggered. The reproduction results are shown in Table~\ref{tab:custom_attack_results}, in which 7 of 10 attacks were fully reproduced, with the same three attacks (DNS/HTTP/STOMP) failing to be observed for the same reasons as in Experiment I (DNS/HTTP due to environmental constraints, TCP STOMP due to dependence on connection responses). This indicates that the behavior of the proposed method functions consistently regardless of sample-specific specification changes.

\begin{table}[t]
\caption{Attack reproduction results for Variant II}
\label{tab:custom_attack_results}
\centering
\footnotesize
\begin{tabular}{cll}
\hline
\textbf{Attack ID} & \textbf{Attack Method} & \textbf{Result} \\
\hline \hline
\texttt{0x91} & GRE Ethernet Flood & Fully reproduced \\
\texttt{0x73} & DNS Flood          & Observation failed (no DNS server) \\
\texttt{0x2d} & HTTP Flood         & Observation failed (no web server) \\
\texttt{0x19} & TCP SYN Flood      & Fully reproduced \\
\texttt{0x5c} & GRE IP Flood       & Fully reproduced \\
\texttt{0x4b} & UDP Generic Flood  & Fully reproduced \\
\texttt{0x82} & TCP ACK Flood      & Fully reproduced \\
\texttt{0x27} & UDP VSE Flood      & Fully reproduced \\
\texttt{0x6a} & UDP Plain Flood    & Fully reproduced \\
\texttt{0x3e} & TCP STOMP Flood    & Partially reproduced (connection attempts only) \\
\hline
\end{tabular}
\end{table}

\subsubsection{Comparison with Experiment I}\label{subsubsec:exp_comparison}

We compare the main results of Experiment I (the original Mirai) and Experiment II (Variant II). Regarding specification extraction accuracy, Experiment I achieved a 100\% match with the ground truth across all 20 items, and Experiment II likewise achieved a complete match across all four primary items. Regarding attack reproduction success rate, both experiments resulted in equivalent outcomes of 7 out of 10 fully reproduced, and the three attacks that failed to be observed (DNS/HTTP/STOMP) were exactly identical in both experiments.

The fact that the number of successful attack reproductions and the failure pattern coincide across both experiments demonstrates that the proposed method functions stably without depending on individual sample specifications or environmental differences. In particular, Variant II showed that the LLM correctly extracted communication specifications that do not exist in the publicly available Mirai source code (Magic Number \texttt{0xCAFEBABE}, encryption key \texttt{0xDEADBEEF}, C2 IP \texttt{192.168.99.99}, and the reassigned attack Vector ID system), demonstrating that the LLM of the proposed method can infer specifications from binary structures without relying on pre-trained data.

\section{Discussion}\label{sec:discussion}

\subsection{Logical Interpretation Capability in Communication Specification Extraction}\label{subsec:exp1_discussion}

The specification extraction accuracy in Experiment I reached \textbf{20/20 items (100\%)}. A particularly noteworthy point is that the LLM correctly recognized the fact that Vector ID \texttt{0x08} is, by definition, a missing entry (Table~\ref{tab:extraction_results}). Because LLMs generate responses based on the probabilistic prediction of tokens, they are prone to hallucinations such as ``filling in the gap with \texttt{0x08} after \texttt{0x07}'' when generating a sequential list. Nevertheless, in the proposed method, the LLM logically interpreted the structure of the \texttt{switch-case} statement implementing attack vector selection during its analysis of the decompiled output, and identified the fact that ``no branch label corresponding to \texttt{0x08} exists'' without hallucination.

This observation suggests that the proposed method is not merely performing pattern matching or keyword extraction, but is rather \textbf{extracting specifications based on a semantic interpretation of the program's control structures}. The fundamental reason why the correspondences between attack IDs (10 types within \texttt{0x00}--\texttt{0x0A} excluding \texttt{0x08}) and each attack function (\texttt{attack\_udp\_generic}, etc.) could be comprehensively and accurately extracted is considered to lie in this logical interpretation capability. The fact that the pseudo-C2 server automatically generated in Step 2 was able to reproduce diverse attacks rests on this precise specification extraction in Step 1.

\subsection{Inferential Capability of the LLM on Binary Structures}\label{subsec:llm_inference}

The results of Experiment II shown in Section~\ref{subsec:experiment2} demonstrated that the LLM of the proposed method can infer specifications from binary structures without relying on pre-trained data. In Variant II, the LLM extracted, in agreement with the ground truth, all of the following items that do not appear in the publicly available Mirai source code: Magic Number \texttt{0xCAFEBABE}, encryption key \texttt{0xDEADBEEF}, C2 IP address \texttt{192.168.99.99}, and the completely reassigned attack Vector ID system. These values are unique to the variant derived from the original Mirai of Experiment I, and are, in principle, unobtainable from public information sources.

In particular, the fact that the LLM was able to accurately follow the reassignment of attack Vector IDs is significant. There was a real possibility that the LLM, influenced by the legacy system (\texttt{0x00}--\texttt{0x0A}) contained in its pre-trained data, would output incorrect correspondences; in fact, however, it completely reconstructed the new system by analyzing the registration logic of the \texttt{attack\_init} function. Combined with the logical interpretation capability discussed in Section~\ref{subsec:exp1_discussion}, this strongly demonstrates that the LLM is performing inference grounded in the actual state of the binary.

This result empirically rules out the fundamental concern that ``the LLM may merely be guessing the correct answers based on prior knowledge.'' Therefore, the proposed method is suggested to be applicable, in principle, even to novel IoT malware samples with previously unseen communication specifications.

\subsection{Importance of the Tool Support Mechanism}\label{subsec:tool_importance}

The initial failure in Magic Number extraction observed in Section~\ref{subsubsec:exp2_extraction} provides an important insight for this study. That is, no matter how sophisticated the inferential capabilities of an LLM are, inference cannot succeed unless information about the analysis target binary is provided to it as context. Because the existing GhidraMCP did not have a function for extracting constant values at specific addresses, the LLM could not access the raw byte sequences in the memory region corresponding to the Magic Number, and the specification extraction did not succeed.

Through the GhidraMCP extension implemented in this study (Section~\ref{subsec:tool_support}), the LLM became able to access the constant values at any address in the binary, and as a result, accurate extraction was achieved for all items. This insight suggests that, in future research on LLM-assisted binary analysis, the essential point lies in optimizing not the ``capability of the LLM alone'' but the ``\textbf{combination of the LLM and information extraction tools}.'' In particular, the question of what kinds of binary information should be provided to the LLM in reverse engineering tasks holds value as an independent research challenge for the future.

\subsection{Limitations of This Study}\label{subsec:limitations}

This study currently has the following limitations, which we explicitly state as challenges for future verification.

First, there are \textbf{constraints on the attack observation environment}. For DNS Flood and HTTP Flood, no attack packets were observed; this is not a control failure of the proposed method but stems from constraints of the experimental environment. These attack methods presuppose, prior to issuing attack packets, the resolution of an external domain name or the completion of a TCP three-way handshake with a web server. Since the closed network of this experiment did not contain a responsive DNS server or web server, it is considered that the bot timed out at the connection-attempt stage before the attack began. The fact that the same three observation-failure attacks (DNS/HTTP/STOMP) coincide exactly between Experiments I and II supports the interpretation that the observation failure stems from the environment side rather than from the method side. Going forward, we expect that integrating a virtual honeypot (a responsive DNS/web server) into the experimental environment will make these attack methods observable as well.

Second, there is the \textbf{limited scope of applicability}. While we demonstrated that the proposed method is effective for analyzing Mirai and its derivative variants, the applicability to other IoT malware families (Gafgyt/BASHLITE, etc.) and to samples with more sophisticated encrypted communication or custom protocols remains unverified. Although Experiment II demonstrated that the LLM of the proposed method can infer specifications from binary structures, whether comparable accuracy is obtained for malware families with different design philosophies requires separate verification.

Third, the \textbf{quantitative evaluation of long-term stability and cost-effectiveness} is insufficient. The experiment in this study was confined to short-term verification centered on 10-second attack observations, and verification of long-term connection maintenance (such as the continuity of keep-alive processing) over hours or days has not been conducted. In addition, regarding the analysis-time reduction effect of the proposed method, a quantitative comparison with conventional manual analysis remains as a challenge for future work.

\section{Conclusion}\label{sec:conclusion}

This paper proposed a pseudo-C2 server automatic generation system that integrates Ghidra with a Large Language Model (LLM), addressing the challenges of short-lived C2 servers and increasing analysis cost in IoT malware dynamic analysis. Experiments on Mirai demonstrated the three contributions of Section~\ref{sec:introduction}: the LLM extracted all \textbf{20 items} of the core protocol elements with 100\% accuracy by semantically interpreting control structures such as \texttt{switch-case} statements; the generated pseudo-C2 server fully reproduced 7 of 10 DDoS attacks (TCP/UDP/GRE) with attack behavior consistent with the original C2; and the resulting analysis platform automates a process that traditionally required advanced reverse engineering skills. Furthermore, applying the method to a customized variant of the publicly available Mirai source code (Experiment II) succeeded end-to-end, demonstrating that the LLM infers specifications from binary structures without relying on pre-trained data.

The significance of this study lies in extending LLM applicability from static analysis assistance to automatic generation of attack infrastructure for dynamic analysis, suggesting applicability to novel IoT malware with previously unseen specifications.

Future work includes: (i) verification on other malware families such as Gafgyt (BASHLITE) to enhance generality; (ii) integrating a virtual honeypot (DNS/web servers) to observe all attack vectors including DNS/HTTP Flood; (iii) verifying long-term stability and quantitatively evaluating analysis-time reduction; and (iv) fine-tuning open-source LLMs (such as Llama) for malware analysis to ensure confidentiality and reduce dependence on commercial LLMs.

\section*{Acknowledgments}
We acknowledge the use of Claude AI assistant (Anthropic) for the English translation of this paper.

\bibliographystyle{unsrtnat}
\bibliography{refs}

\end{document}